# The London-Anderson-Englert-Brout-Higgs-Guralnik-Hagen-Kibble-Weinberg mechanism and Higgs boson reveal the unity and future excitement of physics

Roland E. Allen

*Physics and Astronomy Department, Texas A&M University, College Station, TX USA*

allen@tamu.edu

.

The particle recently discovered by the CMS and ATLAS collaborations at CERN is almost certainly a Higgs boson, fulfilling a quest that can be traced back to three seminal high-energy papers of 1964, but which is intimately connected to ideas in other areas of physics that go back much further. One might oversimplify the history of the features which (i) give mass to the W and Z particles that mediate the weak nuclear interaction, (ii) *effectively* break gauge invariance, (iii) eliminate physically unacceptable Nambu-Goldstone bosons, and (iv) give mass to fermions (like the electron) by collectively calling them the London-Anderson-Englert-Brout-Higgs-Guralnik-Hagen-Kibble-Weinberg mechanism. More important are the implications for the future: a Higgs boson appears to point toward supersymmetry, since new physics is required to protect its mass from enormous quantum corrections, while the discovery of neutrino masses seems to point toward grand unification of the nongravitational forces.

**Introduction**

In 1935, Fritz and Heinz London [1] effectively gave mass to the photon in a superconductor, and thereby provided a macroscopic explanation of the Meissner effect – the expulsion of a magnetic field from a superconductor. From a modern perspective, this mechanism for giving mass to a vector boson can be interpreted as implying an effective breaking of gauge invariance. In 1963, following closely related treatments by himself and others [2], Philip Anderson [3] pointed out another aspect important for the construction of models in particle physics: A would-be zero-mass Nambu-Goldstone boson in a superconductor is effectively eaten by the photon to become a finite-mass longitudinal mode, which appears as a plasmon in a nonrelativistic treatment. The plasma frequency

$$\omega_p = \sqrt{\frac{4\pi n_e e^2}{m_e}} \quad (1.1)$$

can be interpreted as the long-wavelength limit of the frequency of a longitudinal mode which has the same mass as the transverse modes described by (2.6) and (2.8) below. In 1964, realistic models with Lorentz invariance and nonabelian gauge fields were formulated by Englert and Brout [4], Higgs [5,6], and Guralnik, Hagen, and Kibble [7]. The prediction of an observable boson was made by Higgs [6,8] and emphasized by Ellis, Gaillard, and Nanopoulos [9] and others. Finally, Weinberg [10] recognized that a Yukawa interaction with the Higgs field would give masses to fermions like the electron in the fully developed electroweak theory [10,11,12]. So one might collectively call the set of all four essential features the London-Anderson-Englert-Brout-Higgs-Guralnik-Hagen-Kibble-Weinberg (LAEBHGHKW) mechanism for giving masses to fundamental particles. Of course, this list leaves out the critical contributions of many others in the rich history, which has been summarized an enormous number of times in reviews and books, but with the earliest origins outside particle physics usually omitted or de-emphasized.

**Photon mass and breaking of gauge invariance in a superconductor**

Weinberg has stated numerous times that "A superconductor is simply a material in which electromagnetic gauge invariance is spontaneously broken" and has given plausibility arguments why the principal properties of a superconductor should follow from the breaking of gauge invariance [13]. From a modern perspective, this idea originates with the 1935 paper of the London brothers [1], who postulated that

$$\nabla \times \mathbf{j}_s = -\frac{n_s e^2}{mc} \mathbf{B} \qquad (2.1)$$

so that a magnetic field $\mathbf{B}$ induces diamagnetic currents $\mathbf{j}_s$. This equation follows if one assumes that [14]

$$\mathbf{j}_s = -\frac{n_s e^2}{mc} \mathbf{A} \qquad (2.2)$$

where $\mathbf{A}$ is the vector potential. And this last equation is obtained if we assume (in modern nomenclature) an order parameter $\psi_s$ which does not break translational invariance, in the sense that it is uniform in space (just as the Higgs vacuum expectation value $\langle \phi_H \rangle$ is assumed not to break Lorentz/Poincaré invariance): The electric current density is given by

$$\mathbf{j}_s = n_s q \mathbf{v}_s = n_s \frac{q}{m} \mathbf{P}_s = n_s \frac{q}{m}\left(\mathbf{p}_s - \frac{q}{c}\mathbf{A}\right) \qquad (2.3)$$

classically, where $\mathbf{P}_s$ is the mechanical momentum and $\mathbf{p}_s$ is the canonical momentum, or

$$\mathbf{j}_s = \operatorname{Re}\left[\frac{q}{m}\psi_s^*\left(-i\hbar\nabla - \frac{q}{c}\mathbf{A}\right)\psi_s\right] = -n_s \frac{q^2}{mc}\mathbf{A} \quad , \quad n_s = |\psi_s|^2 \qquad (2.4)$$

quantum mechanically, since $\nabla \psi_s = 0$. After the Ginzburg-Landau and BCS theories, and subsequent experimental discoveries, the interpretation is $q = -2e$, with $m$ an effective mass, and with the condensate of Cooper pairs corresponding to the Higgs condensate.

Strictly speaking, (2.2) holds only in a particular gauge (the London gauge), and in an exact treatment the fundamental requirement of gauge invariance still holds both in a superconductor and in high energy physics [2,15] if the ground state or vacuum is included. However, there is an *effective* breaking of gauge invariance – i.e. a breaking of gauge invariance if only the excitations above the ground state or particles above the vacuum are included – which reveals itself in various ways. First, the argument in the paragraph above shows that (2.2) follows if the order parameter is invariant under translations. I.e., this kind of translational invariance in the ground state implies an equation which is manifestly not invariant under a gauge transformation $\mathbf{A} \to \mathbf{A}' = \mathbf{A} + \nabla \Lambda$. (In the present paper we ignore the rich variety of phenomena in condensed matter physics which involve ground states that are not translationally invariant or which are otherwise more complex than the simplest superconductors.) Furthermore, either (2.1) or (2.2) effectively implies a mass for the photon according to the following argument: The two (static) Maxwell equations

$$\nabla \times \mathbf{B} = \frac{4\pi}{c} \mathbf{j}_s \quad , \quad \nabla \cdot \mathbf{B} = 0, \tag{2.5}$$

together with (2.1) and $\nabla \times (\nabla \times \mathbf{B}) = \nabla(\nabla \cdot \mathbf{B}) - \nabla^2 \mathbf{B}$, imply that

$$\nabla^2 \mathbf{B} = m_{ph}^2 \mathbf{B} \quad , \quad m_{ph} = \frac{1}{\lambda_L} \quad , \quad \lambda_L \equiv \sqrt{\frac{mc^2}{4\pi n_s e^2}}. \tag{2.6}$$

Here $m_{ph}$ is the scaled mass, related to the mass $M_{ph}$ in standard units by $m_{ph} = M_{ph} c / \hbar$. In order to obtain (2.6) directly we must replace the original gauge-invariant Maxwell equation (with no external current or time dependence), $\nabla \times \mathbf{B} = 0$, by

$$\nabla \times \mathbf{B} + m_{ph}^2 \mathbf{A} = 0 \tag{2.7}$$

which is again manifestly not gauge-invariant. Since $\mathbf{B} = \nabla \times \mathbf{A}$, this can also be written $\nabla(\nabla \cdot \mathbf{A}) - \nabla^2 \mathbf{A} + m_{ph}^2 \mathbf{A} = 0$. But subjecting this to $\nabla \cdot$ gives a cancellation of the first two terms, so that $\nabla \cdot \mathbf{A} = 0$ and $-\nabla^2 \mathbf{A} + m_{ph}^2 \mathbf{A} = 0$. This is the zero frequency limit of the wave equation for a massive photon

$$\frac{1}{c^2}\frac{\partial^2}{\partial t^2}\mathbf{A} - \nabla^2 \mathbf{A} + m_{ph}^2 \mathbf{A} = 0. \qquad (2.8)$$

At zero frequency, of course, at least one wavevector component must be imaginary, and the Meissner effect follows, as in the last paragraph of this section.

The action which leads to Eq. (2.8) also lacks gauge invariance because of the mass term. The London theory thus already contains 2 of the 4 essential features of the LAEBHGHKW mechanism, with (an effective) spontaneous breaking of gauge symmetry and a mass for the gauge boson, which is in this case the photon.

In the electroweak theory, the vacuum, and thus the vacuum expectation value $\langle \phi_H \rangle$ of the Higgs field, are typically required to have Lorentz/Poincaré invariance. This requirement then leads to the result that gauge invariance is broken. Similarly, when the order parameter $\psi_s$ in a superconductor is required to have translational invariance, (2.4) shows that gauge invariance is broken. The requirement that $\psi_s$ be invariant under translations is less compelling for a superconductor, because $|\psi_s|^2$ rather than $\psi_s$ appears in the photon mass, whereas $\langle \phi_H \rangle$ itself appears in fermion masses according to Eq. (3.18) below. However, $\psi_s$ can be interpreted as the expectation value for an electron-pair field, and it is natural to require that it also be translationally invariant in the ground state. (It should be emphasized that all the reasoning here is for a *ground state* of either the superconductor or the universe.) One might adopt the position that gauge invariance is broken if a more fundamental requirement, translational invariance of the vacuum or ground state, is to be preserved.

As a final connection with the electroweak theory, suppose that a current $\mathbf{j}$ of electron quasiparticles (and quasiholes) is added to the modified Maxwell equation (2.7):

$$\nabla \times \mathbf{B} + m_{ph}^2 \mathbf{A} = \frac{4\pi}{c}\mathbf{j} \ . \tag{2.9}$$

Applying $\nabla \cdot$ gives $\nabla \cdot \mathbf{j} \propto \nabla \cdot \mathbf{A}$, and then by (2.2) $\nabla \cdot (\mathbf{j} + \mathbf{j}_s) = 0$, rather than $\nabla \cdot \mathbf{j} = 0$, so the quasiparticle current is not conserved. Instead, the condensate acts essentially as a reservoir of electron Cooper pairs. This is, of course, another result of the effective breaking of gauge invariance: A conservation law required by a symmetry, according to Noether's theorem, no longer holds when the symmetry is broken. I.e., if the "vacuum" is included, charge is conserved, but charge is not conserved for excitations above the "vacuum", which are described by $\mathbf{j}$. This basic effect is displayed in Andreev reflection, where a negatively-charged electron is reflected as a positively-charged hole. In the same way, the initial conservation laws for the $U(1)$ weak hypercharge $Y$ and the $SU(2)$ weak isospin $T^3$ no longer hold after the Higgs condensate forms, and all that is left is conservation of the electric charge $Q = T^3 + Y$.

For a geometry with a planar boundary at $x = 0$, and variation only in the $x$ direction, the solution of (2.6) is $B = B(0)e^{-x/\lambda_L}$ so the magnetic field falls to zero inside the superconductor with a London penetration depth $\lambda_L$. It is a true demonstration of the unity of physics that this Meissner effect in a superconducting metal and the short range of the weak nuclear force in the universe have the same origin: In each case the vector bosons (photons or W and Z bosons) grow masses because they are coupled to a field which forms a condensate at low temperature, as the metal is cooled in the laboratory or the universe expands and cools after the Big Bang.

**Origin of the masses of fundamental particles**

The mass of an atom or human body arises about 99% from the energy of quarks and gluons moving relativistically inside protons and neutrons, in accordance with $E = mc^2$. The mass of an electron, on the other hand, arises from its Yukawa coupling to the Higgs field. The radius of an electron's orbit in the ground state of a hydrogen atom is

$$r_1 = \frac{\hbar^2}{m_e e^2} \qquad (3.1)$$

and similar results hold for other atoms. So if the mass $m_e$ of an electron were zero there would be no atoms, and the formation of ordinary matter would be impossible without the Higgs condensate.

In the Standard Model of particle physics [16-18], scalar bosons are coupled to the gauge bosons through the covariant derivative $D_\mu = \partial_\mu - ig A_\mu^i t^i$ in the action

$$S_b = \int d^4 x\, \phi_b^\dagger(x) D^\mu D_\mu \phi_b(x). \qquad (3.2)$$

There is thus a term proportional to $\phi_b^\dagger(x) A^{i\mu} A_\mu^j t^i t^j \phi_b(x)$ which has the potential to become a mass term with the form $m^2 A'^{i\mu} A'^i_\mu$ if (i) the scalar boson field $\phi_b$ undergoes condensation, acquiring a nonzero vacuum expectation value, and (ii) the generators $t^i$ behave properly. This happens in the electroweak theory because the remaining action for the Higgs field has the Ginzburg-Landau form $-\mu^2 \phi_b^\dagger \phi_b + \frac{1}{2}\lambda\left(\phi_b^\dagger \phi_b\right)^2$, and $t^i = \sigma^i/2$ in the nonabelian part of $D_\mu$, where the $\sigma^i$ are the Pauli matrices.

More precisely, in the electroweak theory the covariant derivative is

$$D_\mu = \partial_\mu - ig A_\mu^i T^i - ig' B_\mu Y \qquad (3.3)$$

where $T^i$ and $Y$ are respectively the operators for the $SU(2)$ weak isospin and $U(1)$ weak hypercharge. In the representation with weak isospin ½, to which the Higgs field belongs, the generators are $T^i = \frac{1}{2}\sigma^i$, with $i = 1,2,3$, and with the same notation used for operators and their matrix representations. The Higgs field also has weak hypercharge ½, so

$$D_\mu \phi_H = \left(\partial_\mu - igA_\mu^i \frac{\sigma^i}{2} - ig'B_\mu \frac{1}{2}\right)\phi_H \tag{3.4}$$

with

$$\langle \phi_H \rangle = \frac{1}{\sqrt{2}}\begin{pmatrix} 0 \\ v \end{pmatrix} \tag{3.5}$$

after symmetry-breaking, where the treatment here and below is restricted to the unitarity gauge (just as the treatment of a superconductor was restricted to the London gauge). Algebra then gives a term

$$\frac{1}{2}\frac{v^2}{4}\left[g^2\left(A_\mu^1\right)^2 + g^2\left(A_\mu^2\right)^2 + \left(-gA_\mu^3 + g'B_\mu\right)^2\right] \tag{3.6}$$

in the action. This expression can be rewritten in terms of mass and charge eigenstates, which are linear combinations of the original $SU(2)$ and $U(1)$ fields, with the first three mediating the weak nuclear interaction and the fourth being the photon:

$$W_\mu^\pm = \frac{1}{\sqrt{2}}\left(A_\mu^1 \mp iA_\mu^2\right) \qquad \text{with mass } m_W = g\frac{v}{2} \tag{3.7}$$

$$Z_\mu^0 = \frac{1}{\sqrt{g^2 + g'^2}}\left(gA_\mu^3 - g'B_\mu\right) \qquad \text{with mass } m_Z = \sqrt{g^2 + g'^2}\,\frac{v}{2} \tag{3.8}$$

$$A_\mu = \frac{1}{\sqrt{g^2 + g'^2}}\left(g'A_\mu^3 + gB_\mu\right) \qquad \text{with mass } m_A = 0 \;. \tag{3.9}$$

The electric charge operator is defined by

$$Q = T^3 + Y \;. \tag{3.10}$$

With

$$T^{\pm} = T^1 \pm iT^2 = \frac{1}{2}(\sigma^1 \pm i\sigma^2) \qquad (3.11)$$

algebra then gives

$$D_\mu = \partial_\mu - i\frac{g}{\sqrt{2}}(W_\mu^+ T^+ + W_\mu^- T^-) - i\frac{g}{\cos\theta_w}Z_\mu(T^3 - \sin^2\theta_w Q) - ieA_\mu Q \qquad (3.12)$$

where the fundamental electric charge $e$ and weak mixing angle $\theta_w$ are defined by

$$e = g\sin\theta_w \,,\; \sin\theta_w = \frac{g'}{\sqrt{g^2 + g'^2}} \,,\; \cos\theta_w = \frac{g}{\sqrt{g^2 + g'^2}} \qquad (3.13)$$

so that

$$\begin{pmatrix} Z^0 \\ A \end{pmatrix} = \begin{pmatrix} \cos\theta_w & -\sin\theta_w \\ \sin\theta_w & \cos\theta_w \end{pmatrix}\begin{pmatrix} A^3 \\ B \end{pmatrix} \,,\; m_W = m_Z \cos\theta_w \,. \qquad (3.14)$$

A critical feature is that only the 2-component left-handed parts of the fermion fields experience the $SU(2)$ weak interaction. In the Weyl representation the Dirac equation is

$$\begin{pmatrix} -m_f & i\sigma \cdot \partial \\ i\bar{\sigma} \cdot \partial & -m_f \end{pmatrix}\begin{pmatrix} \psi_L \\ \psi_R \end{pmatrix} = 0 \qquad (3.15)$$

with $\sigma^\mu = (1,\sigma)$ and $\bar{\sigma}^\mu = (1,-\sigma)$ in a standard notational convention, so the fermion mass $m_f$ couples left- and right-handed fields (as is required by Lorentz invariance). The 2-component right-handed fermion fields are $SU(2)$ singlets $e_R^-$, $u_R$, $d_R$ (with weak isospin $= 0$) and the left-handed fields are placed into doublets:

$$E_L = \begin{pmatrix} \nu_e \\ e^- \end{pmatrix}_L \text{ and } Q_L = \begin{pmatrix} u \\ d \end{pmatrix}_L, \text{ with } t^3 = \pm\frac{1}{2} \text{ and } y = -\frac{1}{2} \text{ or } y = +\frac{1}{6} \qquad (3.16)$$

where $t^3$ and $y$ are respectively the eigenvalues of $T^3$ and $Y$. Here the upper (or lower) sign corresponds to the upper (or lower) component of each field, and both components of a given doublet have the same weak hypercharge. Recall that $Q = T^3 + Y$, or $q = t^3 + y$, so we get the correct charges $q$ for neutrino $v_e$, electron $e^-$, up quark, and down quark. The Higgs field is also a doublet:

$$\phi_H = \frac{1}{\sqrt{2}} \begin{pmatrix} -i(\phi^1 - i\phi^2) \\ v + (h + i\phi^3) \end{pmatrix}$$

$$\rightarrow \frac{1}{\sqrt{2}} \begin{pmatrix} 0 \\ v \end{pmatrix} \text{ in the ground state}$$

(3.17)

where $h$ represents the massive Higgs boson. The 3 $\phi^i$ are would-be Nambu-Goldstone bosons, which are eaten by the $W^+$, $W^-$, and $Z^0$ vector bosons when they become massive and thus acquire longitudinal polarizations, so that there is no physical Nambu-Goldstone boson.

Notice that a fermion mass term $-m_f \bar{\psi}_L \psi_R - m_f \bar{\psi}_R \psi_L$ in the action again violates gauge invariance, since $\psi_L$ behaves differently from $\psi_R$ under a gauge transformation. The natural way to achieve fermion masses [10] is to postulate Yukawa couplings with the form

$$-\lambda_e \bar{E}_L \cdot \phi_H \, e_R + \text{h.c.}$$

(3.18)

(where h.c. means Hermitian conjugate), which after symmetry breaking becomes

$$-m_e \bar{e}_L e_R + \text{h.c.}, \quad m_e = \frac{1}{\sqrt{2}} \lambda_e v \ .$$

(3.19)

So now the electron and other fermions can have mass, and the weak nuclear force is very short range (mediated by force-carrying particles which have very large masses), all because the Higgs field condensed, acquiring a large vacuum expectation value as the universe cooled after the Big Bang.

**Future physics related to the Higgs in various ways: supersymmetry, grand unification, dark energy, quantum gravity**

The discovery of a scalar boson immediately points to physics beyond the Standard Model, since otherwise radiative corrections should push the mass of this particle up to a ridiculously large value [18]. The most natural candidate for such new physics is supersymmetry (susy), for which there is already indirect experimental evidence, in the sense that the coupling constants of the 3 nongravitational forces are found to converge to a common value (as they are run up to high energy in a grand unified theory) only if the calculation includes susy. In addition, susy predicts a neutralino which is an extremely natural candidate for dark matter. So, instead of acting as an endpoint for physics, and a mere capstone of the Standard Model, the Higgs boson opens the door to a plethora of new particles and effects.

Similarly, the discovery of neutrino masses has opened the door to a more fundamental understanding of forces and matter via grand unification. There are two possibilities for a neutrino mass, either of which is inconsistent with the requirements of the Standard Model: For a Dirac mass, an extra field has to be added for each generation of fermions. For a Majorana mass, lepton number conservation has to be violated. But either or both types of mass are natural with grand unification. At the moment, it is not known whether neutrinos have Majorana masses (in which case a neutrino is its own antiparticle) or Dirac masses or both. This is currently an intense area of research, and any outcome will again involve rich new physics and better understanding of nature.

All of the 4 forces of nature are gauge interactions: The strong nuclear force (quantum chromodynamics) is described by $SU(3)$, and the electromagnetic and weak nuclear forces by $SU(2) \times U(1)$ before symmetry-breaking. At very high energy all of these 3 nongravitational forces are presumably contained in a larger fundamental gauge group, with more than one symmetry-breaking as the universe cooled from a hot Big Bang – for example, with

$$SO(10) \to SU(5) \to SU(3) \times SU(2) \times U(1) \tag{4.1}$$

even before the final symmetry-breaking as the electroweak Higgs condensate formed. And at some energy scale there must be breaking of supersymmetry, which is required not far above 1 TeV if it is to protect the Higgs mass and unify coupling constants, but which certainly does not hold at low energies, since there is, e.g., no selectron with the same mass as the electron. These various high-energy symmetry breakings presumably involve condensation of somewhat Higgs-like fields, and so grand unification is again associated with the Higgs phenomenon.

In Einstein gravity, one has general coordinate transformations, which are described as gauge transformations in contexts like gravitational waves, but under which fermion fields behave as scalars. A local Lorentz transformation, on the other hand, is a true gauge transformation, with fermion fields transforming as spinors. Let $\Sigma^\alpha_\beta$ represent the generators of the Lorentz group (in the spinor representation). The covariant derivative for a fermion is given by

$$D_\mu \psi = \partial_\mu \psi + \frac{1}{2} \omega_\mu^{\alpha\beta} \Sigma_{\alpha\beta} \psi \tag{4.2}$$

where $\omega_\mu^{\alpha\beta}$ is the spin connection, which is the gauge field of the local Lorentz group. The Riemann tensor describing gravity is essentially a gauge field strength given by $\omega_\mu^{\alpha\beta}$ and its derivatives [19].

But despite these similarities with the gauge description of the other forces, the Einstein-Hilbert Lagrangian density for gravity, with the form

$$\mathcal{L}_G = \frac{1}{2} \ell_P^{-2} e\,^{(4)}R \tag{4.3}$$

where $^{(4)}R$ is the curvature scalar, is quite different from the Maxwell-Yang-Mills Lagrangian density for the other forces, with the form

$$\mathcal{L}_g = -\frac{1}{4} g_0^{-2} e F^i_{\mu\nu} F^i_{\rho\sigma} g^{\mu\rho} g^{\nu\sigma} \,, \tag{4.4}$$

where $F^i_{\mu\nu}$ is a gauge curvature, and the coupling to gravity is also quite different. Here $g_0$ is the coupling constant for the fundamental gauge group, $\ell_P^2 = 8\pi G$, where $G$ is the gravitational constant, and $e = \det e^\alpha_\mu = \left(-\det g_{\mu\nu}\right)^{1/2}$, where $e^\alpha_\mu$ is the vierbein and $g_{\mu\nu}$ is the metric tensor. For these reasons it has so far proved impossible to quantize gravity in a realistic theory, despite extremely sophisticated attempts which appear to successfully eliminate the divergences of quantum gravity but which have so far had no success in treating the rest of physics.

Another problem involving gravity is the discovery that most of the energy content of the universe is a mysterious dark energy, which bears a close resemblance to Einstein's cosmological constant [18]. There are actually two mysteries: (i) what is the origin of the dark energy and (ii) why is there not a cosmological constant due to the vacuum energy which is roughly 50 or even 120 orders of magnitude larger than the observed dark energy? This second mystery was taken seriously after it was recognized that the Higgs condensate has an enormous (negative) vacuum energy density, which should show up gravitationally according to conventional physics. So the cosmological constant and dark energy problems are yet again associated with the Higgs phenomenon.

In summary, the discovery of the Higgs boson is a strong reminder of both the essential unity of physics and the 21st Century mysteries that should ultimately lead to a deeper understanding of nature.

## Acknowledgement

I have benefitted from the comments of C. R. Hu.